\documentclass[fleqn,usenatbib]{mnras}

\usepackage{newtxtext,newtxmath}

\usepackage[T1]{fontenc}
\usepackage{ae,aecompl}

\usepackage{graphicx}	
\usepackage{enumerate}
\usepackage{color}
\usepackage{graphicx}
\usepackage{multicol}
\usepackage{savesym}
\usepackage{color}
\usepackage{tipa}
\usepackage{csvsimple}
\usepackage{arydshln}
\usepackage{arydshln}

\title[ULXs in NGC\,1232A]
{X-raying the galaxy pair Arp 41: no collision in NGC\,1232 and three ultraluminous sources in NGC\,1232A}

\author[R. Soria \& M.~W. Pakull]
{Roberto Soria$^{1,2}$ \thanks{Email: rsoria@nao.cas.cn (RS)}, 
Manfred W.~Pakull$^{3}$
\\
$^{1}$College of Astronomy and Space Sciences, University of the Chinese Academy of Sciences, Beijing 100049, China\\
$^{2}$Sydney Institute for Astronomy, School of Physics A28, The University of Sydney, Sydney, NSW 2006, Australia\\
$^{3}$ Universit\'e de Strasbourg, CNRS, Observatoire astronomique, CNRS, UMR 7550,F-67000, Strasbourg, France\\
\\
}

\date{Accepted XXX. Received YYY; in original form ZZZ}

\pubyear{2021}

\begin{document}
\label{firstpage}
\pagerange{\pageref{firstpage}--\pageref{lastpage}}
\maketitle

\begin{abstract}
We studied the apparent galaxy pair NGC\,1232/NGC\,1232A with {\it Chandra}, looking for evidence of interactions and collisions. We report that there is no cloud of diffuse emission in NGC\,1232, contrary to previous claims in the literature. Instead, we find that the small ``companion'' galaxy NGC\,1232A contains three ultraluminous X-ray sources with peak 0.3--10 keV luminosities above $10^{40}$ erg s$^{-1}$ (assuming a cosmological distance of $\approx$93 Mpc for this galaxy). For its mass, morphology, metal abundance and bright ULX population, NGC\,1232A is analogous to the more nearby late-type spiral NGC\,1313.
\end{abstract}


\section{Introduction}
The population properties of X-ray binaries and ultraluminous X-ray sources (ULXs) are a tracer of stellar mass and star formation rate in their host galaxies \citep{lehmer19}. 
In turn, ULXs affect the surrounding interstellar medium with their radiative power and mechanical luminosity, creating X-ray photo-ionized nebulae and/or shock-ionized bubbles \citep{pakull02}. 

As part of this investigation, we are doing a long-term search for ULXs in star-forming galaxies of different morphological types. The galaxy pair NGC\,1232/NGC\,1232A (Arp 41) is a particularly interesting target. The larger galaxy in the pair is the face-on spiral NGC\,1232, of Hubble type SABc, with a Hubble distance $= (21.3 \pm 1.5)$ Mpc (based on its recession speed of $\approx$1450 km s$^{-1}$ with respect to the cosmic background), and a redshift-independent distance $\approx 14.5$ Mpc. See the NASA/IPAC Extragalactic Database (NED\footnote{ https://ned.ipac.caltech.edu.}) for the full reference list of those two distance values; for the redshift-independent distance, we took the median value of 19 different measurements reported in NED, mostly based on the Tully-Fisher relation \citep{tully77}. In this paper, we will use the value of 14.5 Mpc.
A study based on three {\it Chandra} observations from 2008--2010 claimed \citep{garmire13} that there was X-ray evidence for a minor collision, perhaps a dwarf satellite that passed through the disk of NGC\,1232 and shock-ionized the gas around the impact region. Our first objective was to test this widely publicized claim\footnote{See, {\it e.g.}, https://chandra.harvard.edu/photo/2013/ngc1232/}, revisiting the original data and including subsequent {\it Chandra} observations from 2012--2015.

Our second objective was to study the apparent companion, NGC\,1232A. Although at first sight this late-type barred spiral looks like a Magellanic-dwarf satellite of the main galaxy, it has a heliocentric recession speed of $\approx$(6600$\pm$45) km s$^{-1}$ \citep{jones09}, 
which corresponds to a distance modulus of $(34.84 \pm 0.17)$ mag, or a luminosity distance of $(93\pm7)$ Mpc (HyperLEDA database\footnote{http://leda.univ-lyon1.fr}). This is the reason why NGC\,1232 + NGC\,1232A were cited by \cite{arp82} as one of the most striking examples of ``discrepant redshift'' in galaxy pairs. Unfortunately, no reliable redshift-independent distance measurement exists for this galaxy. The peak rotational velocity of the disk in NGC\,1232A 
sometimes listed in galaxy databases ({\it e.g.}, HyperLEDA) is in fact meaningless \citep{fouque90}, because of contamination from the outer spiral arm of NGC\,1232. Therefore, this value cannot be used for Tully-Fisher estimates of distance and baryonic mass.


\begin{figure}
\centering
\includegraphics[width=0.475\textwidth, angle=0]{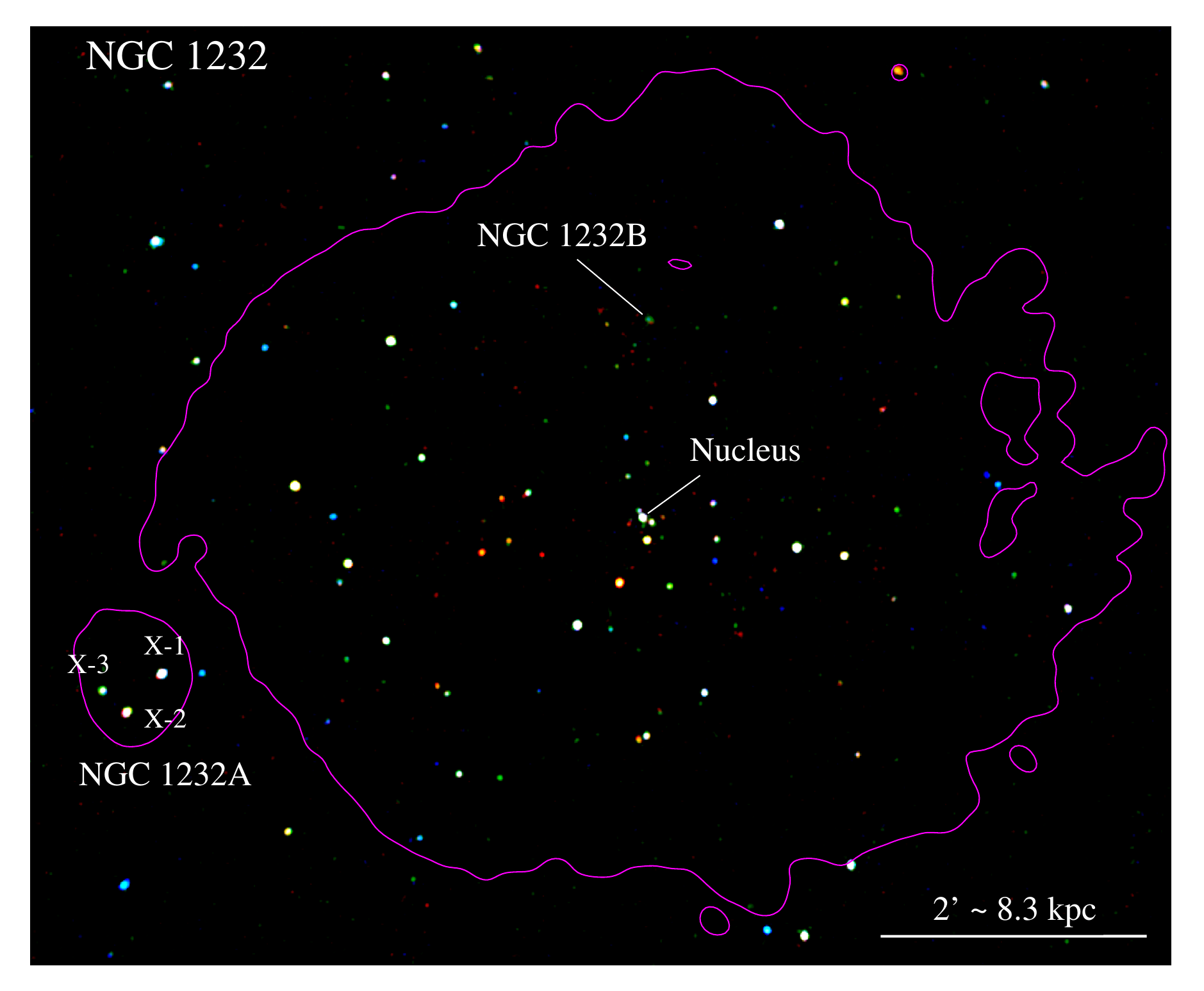}\\[-7pt]
\includegraphics[width=0.475\textwidth, angle=0]{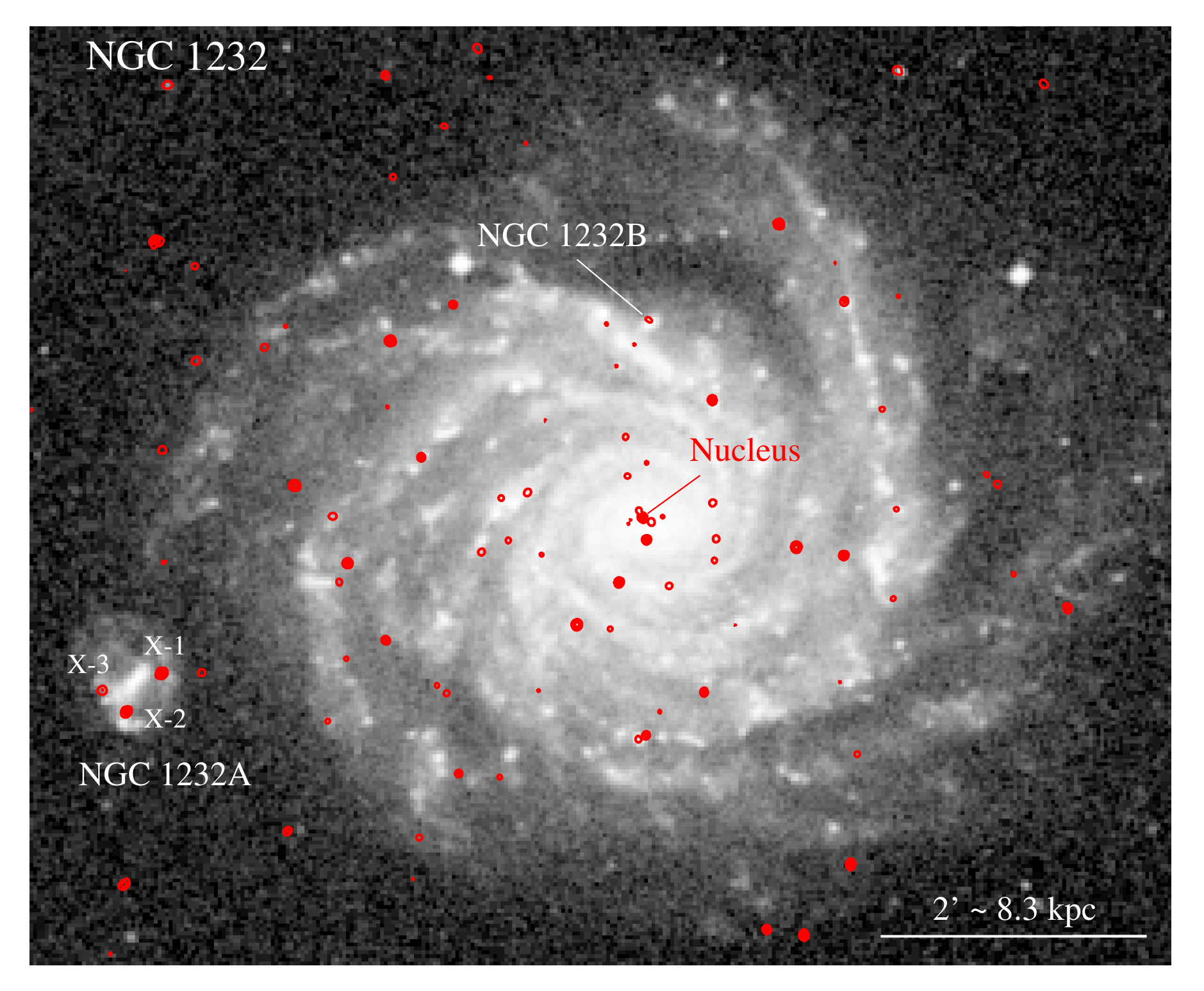}
\vspace{-0.4cm}
 \caption{Top panel: stacked {\it Chandra}/ACIS colour image of NGC\,1232 and NGC\,1232A. Red = 0.3--1.2 keV, green = 1.2--2.6 keV, blue = 2.6--7 keV. North is up and east to the left. The outline of the optical galaxies (roughly corresponding to the D25 isophotes) is overplotted in magenta.  The X-ray nucleus of NGC\,1232, the ULXs in NGC\,1232A, and the background starburst galaxy NGC\,1232B are labelled.
 Bottom panel: Digitized Sky Survey image of the system, with the {\it Chandra} sources overplotted.}
  \label{chandra_sdss}
\end{figure}

\section{Observations and Data Analysis}
The galaxy pair was observed by {\it Chandra}/ACIS on seven occasions between 2008 and 2015: six times with ACIS-I and once with ACIS-S, for a total of 295 ks (Table 1). 
We downloaded the data from the public archives, then reprocessed and analysed them with the Chandra Interactive Analysis of Observations ({\sc ciao}) software version 4.12 \citep{fruscione06}, Calibration Database 4.9.1. In particular, we used the {\it ciao} tasks {\it chandra\_repro} to create new level-2 event files, {\it reproject\_obs} to create stacked images, {\it srcflux} to determine model-independent fluxes, and {\it specextract} to create spectra and associated response and ancillary response files  (both for individual observations and combined for all datasets). 

More specifically, when we extracted the spectra of extended regions (for the study of the diffuse emission in NGC\,1232), we built spatially weighted ancillary response functions ({\it specextract} parameter ``weight = yes'') without a point-source aperture correction ({\it specextract} parameter ``correctpsf = no''). Conversely, when we extracted the spectra of point-like sources (the ULXs in NGC\,1232A), we set ''weight = no'' and ``correctpsf = yes''. Our definition of the extended source and background regions in NGC\,1232 is described in details in Section 3.1. For the point-like sources in NGC\,1232A, we used source extraction circles of $3^{\prime\prime}$ radius, and local background annuli at least 4 times the size of the source regions.

We then used the {\sc ftools} \citep{blackburn95} task {\it grppha} to rebin the spectra to 1 count per bin. Finally,
we modelled the spectra with xspec \citep{arnaud96} version 12.11.0, fitting them with the Cash statistics \citep{cash79}. As a further check of our main results, we rebinned the same spectra to $>$15 counts per bin and re-fitted them with the $\chi^2$ statistics.

\begin{table}
\caption{Summary of the {\it Chandra} observations for NGC\,1232/NGC\,1232A.}
\vspace{-0.4cm}
\begin{center}
\begin{tabular}{lccc}  
\hline \hline\\[-7pt]    
ObsID & Instrument & Obs.~Date &  Exp.~Time (ks)  \\
\hline  \\[-7pt]
10720 & ACIS-I & 2008-11-03 & 47.0 \\ [-2pt]
10798 & ACIS-I & 2008-11-05 & 52.9\\ [-2pt]
12153 & ACIS-I & 2010-09-29 & 50.0\\[-2pt]
12198$^a$ & ACIS-S & 2010-11-30 & 46.5 \\ [-2pt]
14236$^b$ & ACIS-I & 2012-06-13 & 48.9\\ [-2pt]
15391 & ACIS-I & 2013-10-13 & 21.3\\[-2pt]
16486 & ACIS-I & 2013-10-14 & 28.3\\[-2pt]
17463 & ACIS-I & 2015-11-10 & 44.5\\[-2pt]
\hline
\end{tabular} 
\label{tab1}
\end{center}
\vspace{-0.3cm}
\begin{flushleft} 
$^a$: Its field of view covers NGC\,1232A but only half of NGC\,1232.\\
$^b$: Its field of view covers only the north-east corner of NGC\,1232, and does not include NGC\,1232A.\\
\end{flushleft}
\end{table}

\begin{table}
\caption{Diffuse emission in NGC\,1232}
\begin{center}
\begin{tabular}{lcccc}  
\hline \hline\\[-7pt]    
%
$kT_1$ (kev) & $kT_2$ (keV) & Sector & $F_{0.5-2}^a$ & $L_{0.5-2}^b$ \\
%
\hline  \\[-7pt]
$0.30^{+0.23}_{-0.09}$ &  $0.82^{+0.13}_{-0.11}$ & West$^{c}$ &
 $4.24^{+1.13}_{-0.97}$ & $1.15^{+0.31}_{-0.26}$ \\ [3pt]
   &   & North &  $3.27^{+0.94}_{-0.85}$ & $0.88^{+0.25}_{-0.23}$ \\ [3pt]
   &   & East  &  $2.59^{+0.88}_{-0.81}$  & $0.70^{+0.24}_{-0.22}$ \\[3pt]
   &   & South &  $5.01^{+2.23}_{-1.49}$  & $1.38^{+0.65}_{-0.42}$ \\[3pt]
\hline
\end{tabular} 
\label{tab2}
\end{center}
\vspace{-0.3cm}
\begin{flushleft} 
$^a$: observed (absorbed) 0.5--2 keV surface brightness, 
in units of $10^{-15}$ erg cm$^{-2}$ s$^{-1}$ arcmin$^{-2}$;\\
$^b$: unabsorbed 0.5--2 keV surface luminosity, 
in units of $10^{38}$ erg s$^{-1}$ arcmin$^{-2}$, assuming a distance of 14.5 Mpc;\\
$^c$: the alleged hot cloud centre (red circle in \citealt{garmire13}).\\
\end{flushleft}
\end{table}

\begin{figure}
\centering
\includegraphics[width=0.475\textwidth,angle=0]{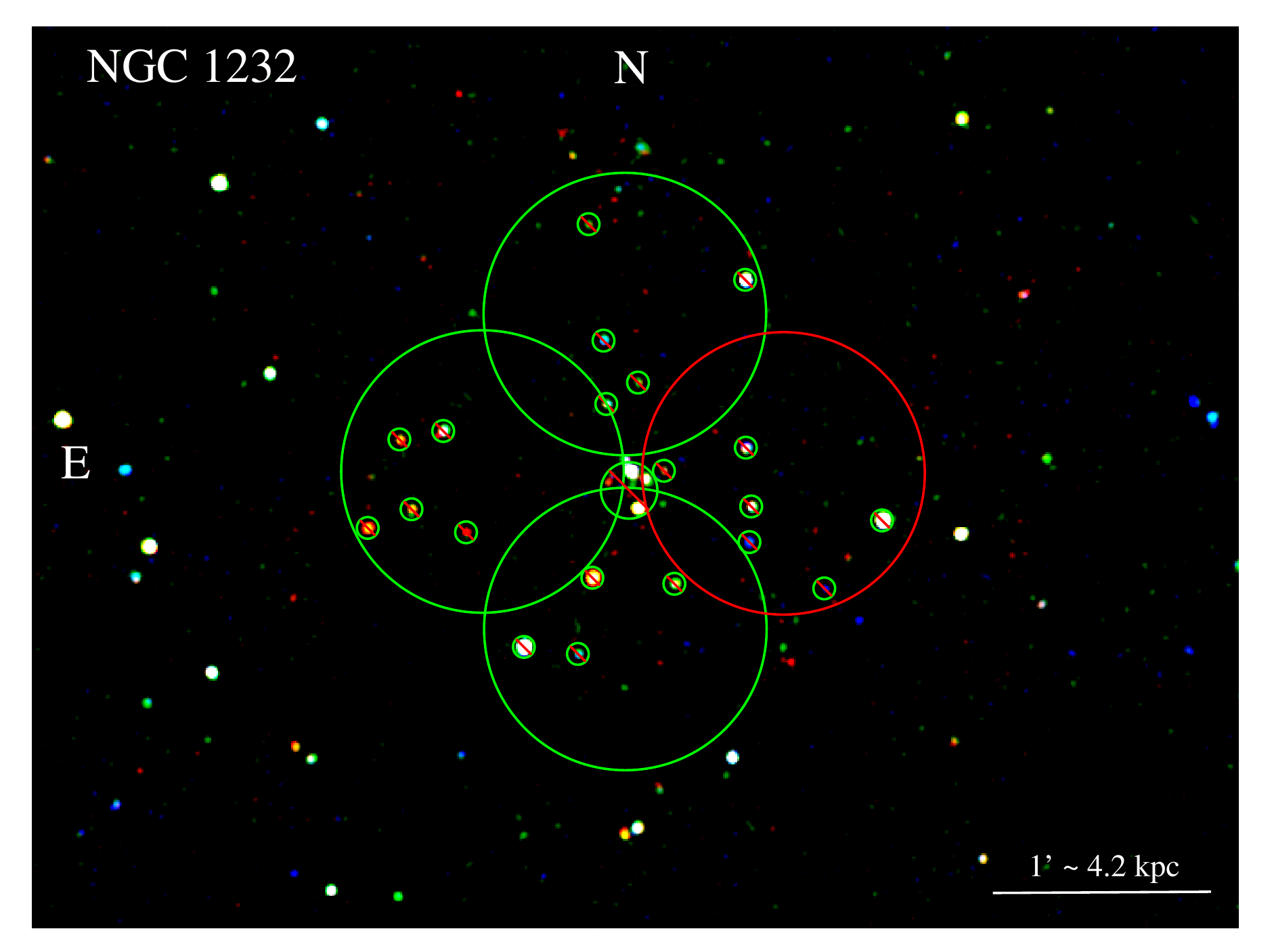}\\[-7pt]
\includegraphics[width=0.475\textwidth, angle=0]{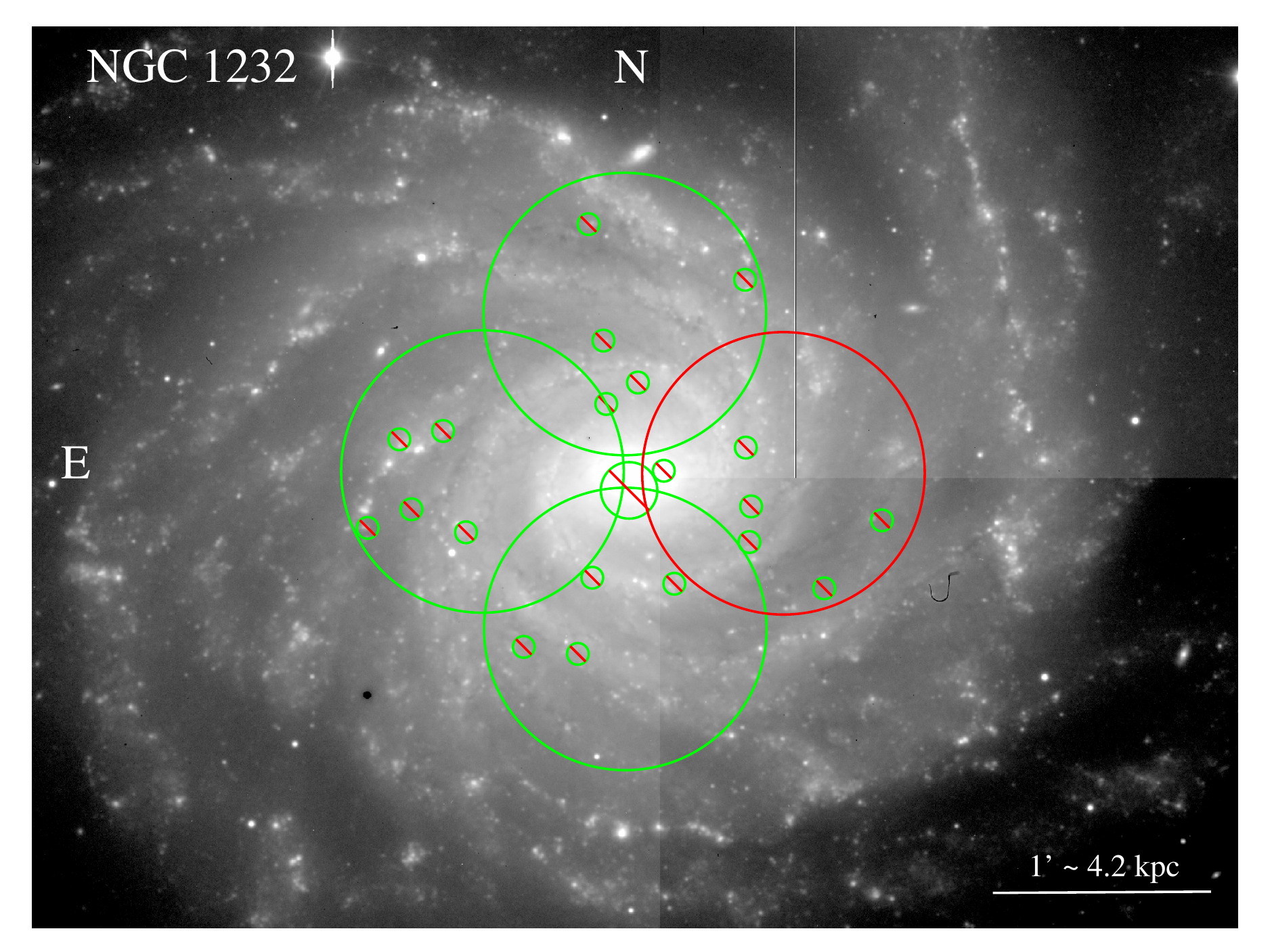}
\vspace{-0.4cm}
 \caption{Top panel: stacked {\it Chandra}/ACIS colour image of the inner disk of NGC\,1232, lightly smoothed with a 3-pixel Gaussian; red = 0.3--1.2 keV, green = 1.2--2.6 keV, blue = 2.6--7 keV. The red circle west of the nucleus corresponds to the extraction region (red circle) in Fig.~2 of Garmire (2013), with a radius of $40^{\prime\prime}$. We positioned an identical circular extraction region at three other locations around the nucleus (north, east and south). We excluded the point sources from our extraction regions (small excluded circles with radii of $3^{\prime\prime}$, plus an excluded circle with a radius of $8^{\prime\prime}$ around the complex nuclear region), because we are looking for diffuse thermal plasma emission. 
 Bottom panel: $R$-band image from VLT/FORS2 taken on 1999 November 11 (source: ESO public archives), with the {\it Chandra} extraction regions overplotted, for direct comparison with Fig.~2 of Garmire (2013). [This VLT image is available in FITS format but unfortunately does not cover NGC\,1232A].}
  \label{chandra_sdss}
\end{figure}

\begin{figure}
\hspace{-0.5cm}
\includegraphics[height=0.49\textwidth, angle=270]{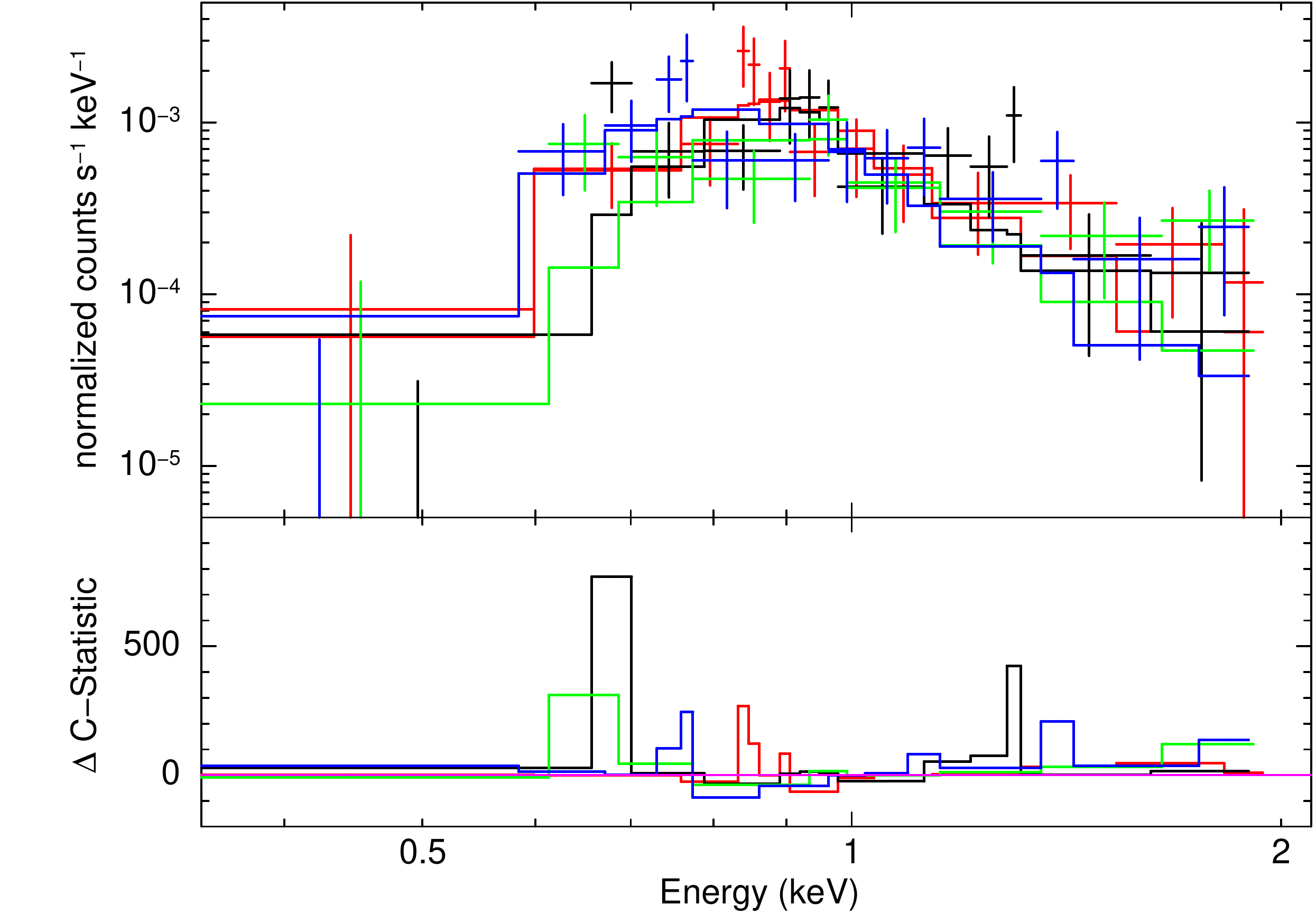}
 \caption{Stacked {\it Chandra}/ACIS-I spectra for each of the four inner-disk regions illustrated in Figure 2, and corresponding C-stat residuals. Red datapoints and fitting model are for the western region (corresponding to the red circle in \citealt{garmire13}); black for the northern region; green for the eastern region; blue for the southern region. There is no significant background-subtracted emission above $\approx$2 keV. The four spectra have been fitted simultaneously with a two-temperature {\it apec} model, with locked values of the two temperatures, free normalization, and only line-of-sight Galactic $N_{\rm}$. See Table 2 for the fit parameters. For plotting purposes, the datapoints have been rebinned to a minimum signal-to-noise ratio $>$2 and maximum number of grouped channels $=$ 20.}
  \label{flux_lc}
\end{figure}

\section{Main Results}

\subsection{No evidence of galaxy collisions in NGC\,1232}
We analyzed the stacked ACIS-I images, looking for an excess of diffuse hot plasma emission to the west and to the north of the nucleus, as claimed by \cite{garmire13}. Neither the distribution of count rates or the fluxes measured with {\it srcflux} suggest any such excess. However, we do notice that our stacked images contain many more point sources (obviously, because they are deeper than the images available to \citealt{garmire13}). Several well-resolved point-like sources were just below the signal-to-noise threshold of the adaptive-smoothing task {\it csmooth} used by \cite{garmire13}, based on the first three {\it Chandra} observations alone. As a result, we suspect that {\it csmooth} treated the photons associated to those faint sources as diffuse emission, and smeared them into a large cloud, whose size is a function of the smoothing-scale parameter. 

To investigate the matter in more details, we reproduced the circular extraction region used by \cite{garmire13} (red circle in our Figure 2, corresponding to the red circle in their Fig.~2, with a radius of 40$^{\prime\prime}$). However, in our case, we removed several point sources inside the large circle, because we are looking only for the diffuse hot gas. The detection threshold for our identification of point-like sources was 12 net counts (0.3--7 keV band) in the stacked ACIS-I image; this corresponds to a 0.3--10 keV luminosity of $\approx$2 $\times 10^{37}$ erg s$^{-1}$ for a power-law photon index $\Gamma = 1.7$. There are about 150 ``diffuse'' net counts inside the red circle, and an additional 190 counts in point-like sources that we have excluded but which were included in the spectrum of \cite{garmire13}.

We extracted background-subtracted spectra (with their corresponding response and ancillary response files) from each ACIS-I exposure and combined them, for a total exposure time of 243 ks. For each exposure, the background region was carefully chosen in the same ACIS-I chip, at the outer edge or outside the star-forming disk of the galaxy (to minimize the possible presence of diffuse hot gas in the background region), and with an area of at least three times the source region. Point-like sources were also excluded from the background regions. We then repeated the same procedure, placing the same sized extraction region at different locations around the nucleus (north, east and south). In each case, we also subtracted the point-like sources inside the extraction circle (Figure 2). All four sectors have a similar number of diffuse net counts (between $\approx$125--150). Finally, we fitted the four stacked spectra (west, north, east and south locations) simultaneously in {\sc xspec}. After initially fitting the spectra over the whole 0.3--7 keV band, we realized that there is no significant emission (no net counts) above the background level at energies $>$2 keV; thus, we limited our spectral fits to the 0.5--2.0 keV range. The inclusion of the energy channels $>$2 keV has only the effect of increasing the noise but does not change the best-fitting values. If the results of \cite{garmire13} are correct, we expect a strong excess of diffuse emission in the northern and western circles compared with the other two.

For our spectral fitting, we drew on our experience of diffuse emission in galaxies, and assumed a two-temperature thermal plasma model ({\it phabs} $\times$ ({\it apec} $+$ {\it apec}) in {\sc xspec}). The temperatures were free parameters but were locked across the four spectra. Not unexpectedly, they converge to values of $\approx$0.3 keV and $\approx$0.8 keV, which are typical of hot gas in normal star-forming galaxies \citep{mineo12b,lehmer15}. The normalizations of the two plasma components were left free for each spectrum. The intrinsic column density (in addition to the line-of-sight component) was fixed to zero. More realistically, the diffuse emission may see an additional absorption of $\sim$10$^{21}$ cm$^{-2}$ through the halo and part of the disk; however, ACIS-I is not very sensitive at soft X-ray energies: this, combined with the low number of counts, makes it impossible to determine the intrinsic $N_{\rm H}$ accurately. 

To ascertain whether our modelling result was robust, we repeated it in several different ways. First, we fitted the four spectra with the Cash statistics (C-stat $= 485.1/422$). Then, we rebinned the data to 15 counts per bin and refitted it with $\chi^2$ statistics ($\chi^2_{\nu} = 125.6/100$). Then, we left the intrinsic absorption as a free parameter: the best-fitting value increases to $N_{\rm {H,int}} \sim 3 \times 10^{21}$ cm$^{-2}$, but remains consistent with 0 within the 90\% confidence limit. We tried unlocking the temperatures between the four spectra, but found that the two sets of values still cluster around 0.3 and 0.8 keV. We also tried replacing the higher-temperature {\it apec} component with a power-law of photon index 1.7, but found that the emission between 0.5--2 keV is still mostly due to the thermal plasma component, with only minor changes to flux and luminosity. 

We conclude that there is significant thermal plasma emission in each of the four regions, in the 0.5--2 keV band (Figure 3). Above 2 keV, we expect an additional power-law component from unresolved point-like sources fainter than $\sim$10$^{37}$ erg s$^{-1}$, but such component is not significant above the background noise in our spectra, and its presence or absence does not substantially affect our estimate of the thermal plasma luminosity at 0.5--2 keV. Our main result (Table 2) is that the flux and luminosity of the diffuse emission in the four quadrants is similar within a factor of 2 (or even less, if we take into account their error bars): an unremarkable variability in a normal galaxy, considering that spiral arms, gas and stellar distributions are never perfectly symmetric. The luminosity of the western sector is not particularly high or remarkable compared with the other three. In fact, the average luminosity of western plus northern sectors (the alleged hot cloud) is identical to the average luminosity of eastern plus southern sectors. By comparison with the flux in the western sector given by \cite{garmire13}, we infer a 0.5--2 keV diffuse, unabsorbed flux of $\approx$4.6 $\times 10^{-15}$ erg cm$^{-2}$ s$^{-1}$ arcmin$^{-2}$, as opposed to their reported value of $\approx$7 $\times 10^{-15}$ erg cm$^{-2}$ s$^{-1}$ arcmin$^{-2}$ (for the same choice of $N_{\rm H}$). Our lower value, as we mentioned earlier, is because we have not included a few point-like sources whose flux was instead included in the measurement of \cite{garmire13}.

The overall hot gas luminosity of NGC\,1232 is also not at all remarkable. Based on the measured X-ray luminosity of the inner disk, we estimate a total hot gas luminosity of $\sim$1--2 $\times 10^{39}$ erg s$^{-1}$ from the whole star-forming disk (radius of $\approx$2$^{\prime}$), at the assumed Tully-Fisher distance of 14.5 Mpc, or a factor of two higher if we assume a cosmological distance of 21 Mpc. This is indeed the range of luminosities expected \citep{mineo12b} for a galaxy with the star formation rate of NGC\,1232 ($\approx$0.4--0.9 $M_{\odot}$ yr$^{-1}$, depending on the assumed distance: \citealt{araujo18}).

In summary, there is no evidence that the diffuse hot gas distribution in NGC\,1232 is in any way modified or enhanced by a recent minor collision. A more thorough discussion on the X-ray properties and luminosity function of the point sources in NGC\,1232 is beyond the scope of this paper.


\begin{table}
\caption{Coordinates of the three ULXs in NGC\,1232A}
\vspace{-0.1cm}
\begin{center}
\begin{tabular}{lcc}  
\hline \hline\\[-8pt]    
Source & R.A.(J2000) & Dec(J2000) \\
\hline  \\[-8pt]
X-1 & 03:10:01.01 & $-$20:35:53.8\\ [-2pt]
X-2 & 03:10:02.14 & $-$20:36:11.1\\ [-2pt]
X-3 & 03:10:02.92 & $-$20:36:01.3\\ [-2pt]
\hline\\[-5pt]
\end{tabular} 
\label{tab3}
\end{center}
\end{table}

\subsection{Surprisingly luminous ULXs in NGC\,1232A}
There are three bright point-like sources inside the D25 extent
of the small, neighbouring galaxy NGC\,1232A. Here, we label them for simplicity X-1, X-2 and X-3 in order of decreasing average flux from the stacked {\it Chandra} image. Their coordinates are listed in Table 3. 
We determined the coordinates of those three sources from a stacked ACIS image of the 7 observations covering NGC\,1232A; we used the centroiding task {\it dmstat} in {\sc ciao}. The 90\% uncertainty radius of ACIS-I observations taken around 2008--2015 is $\approx$0$^{\prime\prime}$8\footnote{https://cxc.harvard.edu/cal/ASPECT/celmon/}. By taking an average of 7 observations, we effectively reduced this uncertainty to $\approx$0$^{\prime\prime}$3. Unfortunately there are no sources with both {\it Gaia} and {\it Chandra} detections within a few arcmin of NGC\,1232A. A few X-ray sources at the outskirts of NGC\,1232, which appear to have a point-like optical counterpart, confirm that the X-ray astrometry of the stacked image is accurate within $\approx$0$^{\prime\prime}$3.

The probability that the faintest source is a background AGN projected by chance inside the D25 of the galaxy is $\approx$0.15 \citep{luo17}. The probability that all three sources are background AGN is $<$10$^{-3}$. We cannot completely rule out that one or more of those three sources belong to NGC\,1232, but there is no evidence that the outer spiral arm of the main galaxy extends in front of NGC\,1232A.
Thus, the simplest explanation is that all three sources belong to the latter galaxy. More specifically, they appear to be associated with its spiral arms rather than the bar or the nuclear region (the nucleus of NGC\,1232A is not detected). 

Unfortunately, there are no high-quality images of NGC\,1232A available in public data archives. By far the best image of the galaxy pair was taken with FORS1 on the Very Large Telescope (VLT) in 1998 September, during the calibration phase of that instrument; it was published as a European Southern Observatory (ESO) press release\footnote{https://www.eso.org/public/images/eso9845d}.
However, the original data for that observation were not archived and have been lost (Mariya Lyubenova, priv.comm.; Nathalie Fourniol, priv.comm.) Nonetheless, we used the public-outreach JPG image from ESO together with a FITS image from the Digitized Sky Survey to illustrate an indicative position of the three ULXs (Figure 4).

\begin{figure}
\centering
\includegraphics[width=0.47\textwidth, angle=0]{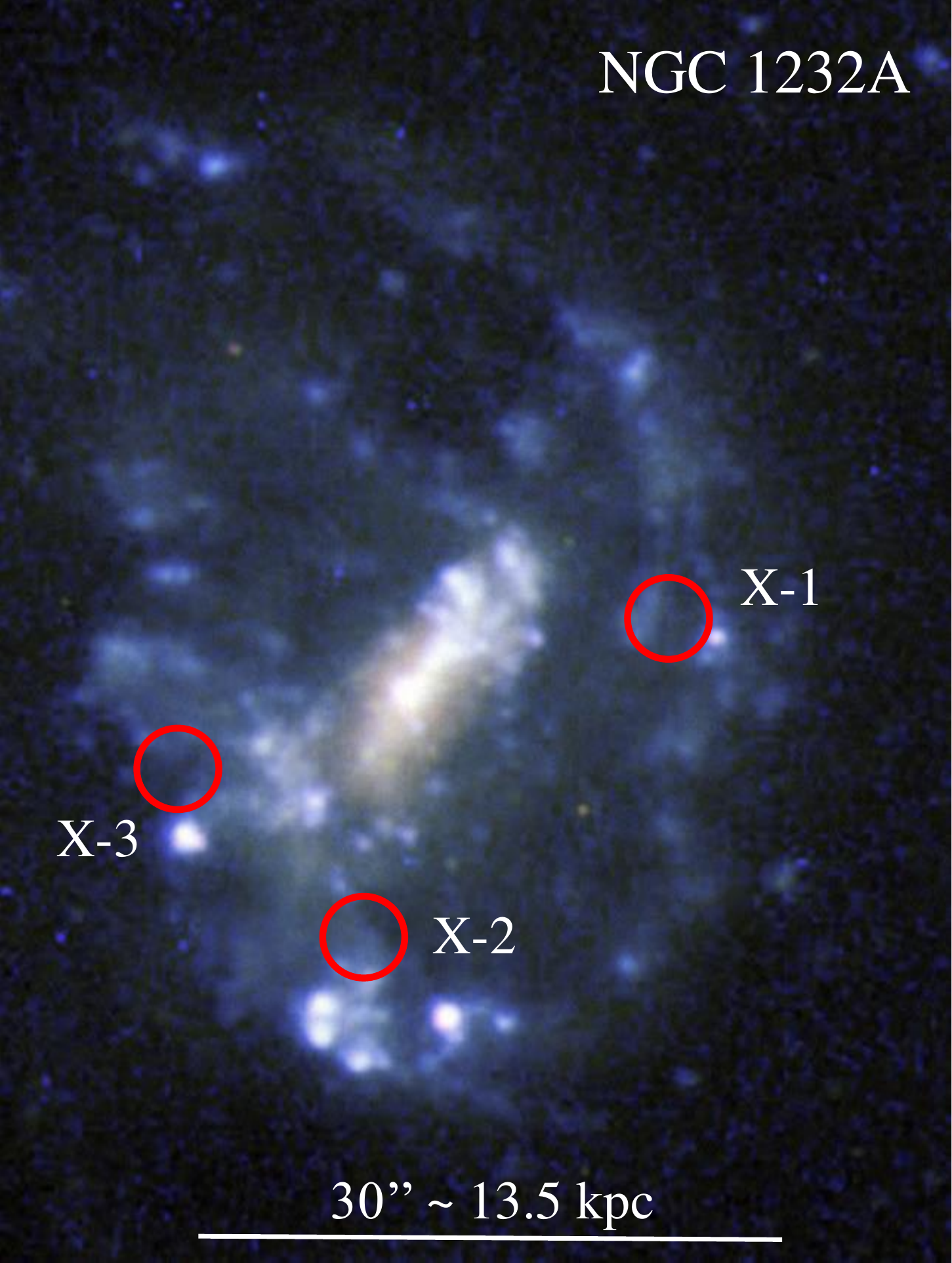}
 \caption{VLT/FORS1 image of NGC\,1232A. Red = $R$ filter; green = $B$ filter; blue = $U$ filter. North is up, east to the left. 
 The approximate 1-$\sigma$ positions for the three ULXs (red circles) was estimated by eye, based on the much lower resolution Digitized Sky Survey image.}
  \label{vlt_fors1}
\end{figure}

We determined the model-independent fluxes of the sources in each observation, to study their long-term variability properties (Figure 5). The observed fluxes
vary by a factor of a few between the seven epochs (X-2 and X-3 being more variable than X-1). This level of irregular variability is a common feature of persistently active ULXs \citep{weng18}.
Then, we modelled the combined spectra for each source.

\begin{figure}
\hspace{-0.5cm}
\includegraphics[height=0.49\textwidth, angle=270]{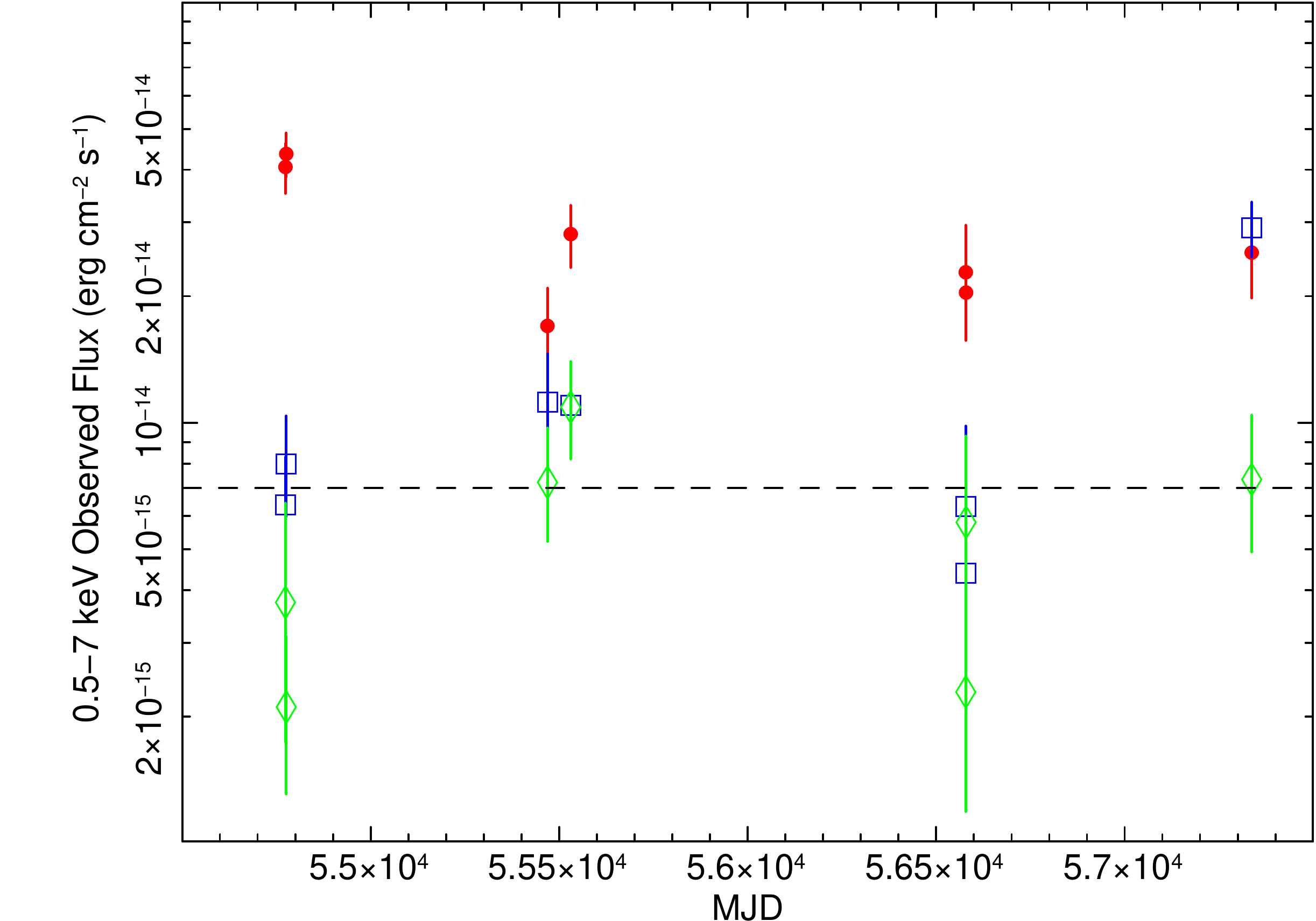}
 \caption{Long-term observed light-curve of the three ULXs over the seven {\it Chandra} observations (red circles = X-1, blue squares = X-2, green diamonds = X-3). The ``observed fluxes'' are the model-independent, background-subtracted 0.5--7 keV values computed with {\it srcflux}. The horizontal dashed line marks a reference 0.3--10 keV luminosity of $\approx$10$^{40}$ erg s$^{-1}$ corresponding to an absorbed 0.5--7 flux of $\approx$ 7 $\times 10^{-15}$ erg s$^{-1}$, for a photon index $\Gamma = 1.7$ and only line-of-sight column density (hence, it is a lower limit for the luminosity if we also account for intrinsic absorption).}
  \label{flux_lc}
\end{figure}

\begin{table}
\caption{Best-fitting parameters of the combined {\it Chandra}/ACIS spectrum of X-1, fitted with the Cash statistics. The Galactic column density is fixed at $N_{\rm {H,Gal}} = 2.1 \times 10^{20}$ cm$^{-2}$.} 
\vspace{-0.3cm}
\begin{center}  
\begin{tabular}{lc} 
 \hline 
\hline \\[-7pt]
  Model Parameters      &      Values \\
\hline\\[-8pt]
\multicolumn{2}{c}{{\it tbabs} $\times$ {\it tbabs} $\times$ {\it po}}\\
\hline\\[-8pt]
   $N_{\rm {H,int}}$   ($10^{20}$ cm$^{-2}$)   &  $ 11.6^{+11.7}_{-10.2}$ \\[4pt]
   $\Gamma$      &  $ 1.32^{+0.18}_{-0.18}$       \\[4pt] 
   $N_{\rm {po}}^a $ 
              &  $ 4.8^{+1.1}_{-0.9}$ \\[4pt]
   C-stat/dof     &      $229.1/277$ (0.83) \\[4pt]
   $f_{0.5-7}$ ($10 ^{-14}$ erg cm$^{-2}$ s$^{-1}$)$^b$ 
       & $ 3.29^{+0.28}_{-0.27}$\\[4pt]
   $L_{0.3-10}$ ($10 ^{40}$ erg  s$^{-1}$)$^c$ 
      & $ 5.08^{+0.47}_{-0.42}$\\[2pt]
   \hline\\[-8pt]
\multicolumn{2}{c}{{\it tbabs} $\times$ {\it tbabs} $\times$ {\it diskbb}}\\
\hline\\[-8pt]
$N_{\rm {H,int}}$   ($10^{20}$ cm$^{-2}$)   &  $ <5.5$  \\[4pt]
   $kT_{\rm in}$ (keV)     &  $ 2.17^{+0.42}_{-0.30}$\\[4pt] 
   $N_{\rm {dbb}} $  ($10^{-4}$ km$^2$)$^d$ &  $ 0.91^{+0.55}_{-0.39}$\\[4pt]
   $R_{\rm {in}}\sqrt{\cos \theta} $  (km)$^e$ & $108^{+29}_{-26}$\\[4pt]
   C-stat/dof     &   $229.4/277$ (0.83)\\[4pt]
   $f_{0.5-7}$ ($10 ^{-14}$ erg cm$^{-2}$ s$^{-1}$)$^b$ 
       & $ 3.23^{+0.27}_{-0.26}$\\[4pt]
   $L_{0.3-10}$ ($10 ^{39}$ erg  s$^{-1}$)$^c$ 
      & $ 4.07^{+0.50}_{-0.41}$\\[2pt]
\hline 
\vspace{-0.5cm}
\end{tabular}
\end{center}
\begin{flushleft} 
$^a$: units of $10^{-6}$ photons keV$^{-1}$ cm$^{-2}$ s$^{-1}$ at 1 keV.\\
$^b$: observed flux in the 0.5--7 keV band\\
$^c$: de-absorbed luminosity in the 0.3--10 keV band, defined as $4\pi d^2$ times the de-absorbed 0.3--10 keV model flux.\\
$^d$: $N_{\rm {dbb}} = (r_{\rm{in}}/d_{10})^2 \cos \theta$, where $r_{\rm{in}}$ is the ``apparent'' inner disk radius in km, $d_{10}$ the distance to the galaxy in units of 10 kpc (here, $d_{10} = 9300$), and $\theta$ is our viewing angle. The total luminosity of the disk is $L \approx 4\pi r_{\rm {in}}^2 \, \sigma T_{\rm{in}}^4$ \citep{makishima86}\\
$^e$: the ``true'' inner-disk radius $R_{\rm{in}}$ is defined as $R_{\rm {in}} \approx 1.19 r_{\rm in}$ for a standard disk \citep{kubota98}, and $r_{\rm {in}}$ was defined in Table note d.\\
\end{flushleft}
\end{table}

\begin{table}
\caption{As in Table 4, for the combined spectrum of X-2.} 
\vspace{-0.2cm}
\begin{center}  
\begin{tabular}{lc} 
 \hline 
\hline \\[-8pt]
  Model Parameters      &      Values \\
\hline\\[-8pt]
\multicolumn{2}{c}{{\it tbabs} $\times$ {\it tbabs} $\times$ ({\it po} $+$ {\it apec})}\\
\hline\\[-8pt]
   $N_{\rm {H,int}}$   ($10^{20}$ cm$^{-2}$)   &  $ 14.3^{+17.0}_{-14.0}$ \\[4pt]
   $\Gamma$      &  $ 1.76^{+0.40}_{-0.44}$       \\[4pt] 
   $N_{\rm {po}}$ 
              &  $ 2.8^{+1.6}_{-1.5}$ \\[4pt]
   $kT_{\rm apec}$ (keV)     &  $ 1.23^{+1.30}_{-0.35}$\\[4pt] 
   $N_{\rm {apec}} $  ($10^{-6}$)$^a$ &  $ 1.7^{+4.3}_{-1.3}$\\[4pt]
   C-stat/dof     &      $135.7/168$ (0.81) \\[4pt]
   $f_{0.5-7}$ ($10^{-14}$ erg cm$^{-2}$ s$^{-1}$) 
       & $ 1.40^{+0.18}_{-0.17}$\\[4pt]
   $L_{0.3-10}$ ($10^{40}$ erg  s$^{-1}$)
      & $ 1.69^{+0.30}_{-0.23}$\\[2pt]
   \hline\\[-7pt]
\multicolumn{2}{c}{{\it tbabs} $\times$ {\it tbabs} $\times$ ({\it diskbb}$+$ {\it apec})}\\
\hline\\[-8pt]
$N_{\rm {H,int}}$   ($10^{20}$ cm$^{-2}$)   &  $<11.2$  \\[4pt]
   $kT_{\rm in}$ (keV)     &  $ 1.50^{+0.70}_{-0.33}$\\[4pt] 
   $N_{\rm {dbb}} $  ($10^{-4}$ km$^2$) &  $ 1.22^{+2.01}_{-0.95}$\\[4pt]
   $R_{\rm {in}}\sqrt{\cos \theta} $  (km) & $125^{+88}_{-67}$\\[4pt]
   $kT_{\rm apec}$ (keV)     &  $ 1.25^{+0.59}_{-0.26}$\\[4pt] 
   $N_{\rm {apec}} $  ($10^{-6}$) &  $ 1.9^{+2.8}_{-1.1}$\\[4pt]
   C-stat/dof     &   $135.4/168$ (0.81)\\[4pt]
   $f_{0.5-7}$ ($10^{-14}$ erg cm$^{-2}$ s$^{-1}$) 
       & $ 1.36^{+0.18}_{-0.17}$\\[4pt]
   $L_{0.3-10}$ ($10^{40}$ erg  s$^{-1}$) 
      & $1.59^{+0.28}_{-0.20}$\\[2pt]
      \hline
\vspace{-0.5cm}
\end{tabular}
\end{center}
\begin{flushleft} 
$^a$: units of $10^{-14}/\{4\pi d^2\}\,\int n_en_{\rm H} \, dV$, where $d$ is the luminosity distance in cm, and $n_e$ and $n_H$ are the electron and H densities in cm$^{-3}$.\\
\end{flushleft}
\end{table}

\subsubsection{X-1: a hard power-law spectrum}
The spectrum of X-1 is best-fitted (C-stat $= 229.1/277$) by a power-law model (no significant curvature), with a rather hard photon index $\Gamma = 1.3 \pm 0.2$ (Table 4, Figure 6), which classifies it in the hard ultraluminous regime \citep{sutton13}. In other words, the spectral luminosity $E\,(dL/dE)$ increases with energy, in the {\it Chandra} band. A disk-blackbody spectrum provides a statistically identical fit (C-stat $= 229.4/277$) but the best-fitting temperature is $\approx$2.2 keV. 
This is too hot \citep{kubota04} and the corresponding luminosity too much above Eddington to be consistent with a standard disk around a stellar-mass compact object; if the spectrum is thermal, it is more likely to be in the slim disk regime \citep{watarai01}.
At 93 Mpc, the average unabsorbed 0.3--10 keV luminosity is $L_{\rm X} \approx$ 4--5 $\times 10^{40}$ erg s$^{-1}$, but as high as $\approx$7 $\times 10^{40}$ erg s$^{-1}$ in both the observations from 2008.

\subsubsection{X-2: a softer ULX with possible thermal plasma emission}
The average spectrum of X-2 is softer than that of X-1. A simple power-law model (C-stat $= 140.5/170$) requires $\Gamma = 2.1 \pm 0.3$: an approximately flat value of the spectral luminosity $E\,(dL/dE)$, so that it is not obvious which of the three ultraluminous regimes defined by \cite{sutton13} it belongs to. 
The fit is marginally improved (C-stat $= 135.7/168$) by the addition of a thermal plasma component at a temperature of $\approx 1.2$ keV; in our {\sc xspec} analysis, we used the {\it apec} model (Table 5). When a thermal plasma component is present, the best-fitting power-law photon index is $\Gamma = 1.8 \pm 0.4$.
As a further check on the significance of the thermal plasma component, we refitted the same spectrum, regrouped to $>$15 counts per bin, with the $\chi^2$ statistics. We find the same best-fitting values. Without the {\it apec} component, we obtain $\chi^2_{\nu} = 16.73/18$; with the {\it apec} component, 
$\chi^2_{\nu} = 12.43/16$. 
Alternatively, we fitted the spectrum with a disk-blackbody model without thermal plasma emission (C-stat $= 148.55/170$; $\chi^2_{\nu} = 23.37/18$) and with thermal plasma emission (Figure 7; C-stat $=135.36/168$; $\chi^2_{\nu} = 12.53/16$). 

Although the improvement provided by the {\it apec} component is statistically small (particularly in the case of the power-law model), adding this component does visually allow the model to do a better job at reproducing the slightly peaked nature of the spectrum around 1 keV (Figure 7), and similar emission has been seen in other ULX systems ({\it e.g.}, \citealt{middleton15}). We therefore report the results both with and without the APEC component, for completeness. The contribution of thermal plasma emission in other ULXs is generally linked to the presence of strong outflows \citep{pinto16,pinto17}, so we think it worthwhile to report this possibility for X-2.


The average 0.3--10 keV unabsorbed luminosity is $L_{\rm X} \approx 1.6 \times 10^{40}$ erg s$^{-1}$ (Table 5), of which $\approx$3 $\times 10^{39}$ erg s$^{-1}$ may be in the thermal plasma component. However, in ObsID 17463, X-2 reached $L_{\rm X} = 5 \times 10^{40}$ erg s$^{-1}$.

\subsubsection{X-3: a lightly curved spectrum}
The stacked spectrum of X-3 has an acceptable fit (Table 6) with a power-law model (C-stat $= 81.8/111$, or $\chi^2_{\nu} = 10.09/7$) with $\Gamma = 2.1 \pm 0.5$ (again, a flat spectral luminosity). It has a marginally better fit (C-stat $= 77.4/111$; $\chi^2_{\nu} = 6.95/7$) with a curved model such as {\it diskbb} (Figure 8), with $kT_{\rm in} \approx 1.2$ keV and inner disk radius $R_{\rm{in}} (\cos \theta)^{1/2} \approx 170$ km. 
The average 0.3--10 keV luminosity is $L_{\rm X} = 0.8 \times 10^{40}$ erg s$^{-1}$ for the disk-blackbody model or $L_{\rm X} = 1.5 \times 10^{40}$ erg s$^{-1}$ for the power-law model (which requires a higher intrinsic column density). 

\begin{figure}
\hspace{-0.5cm}
\includegraphics[height=0.482\textwidth, angle=270]{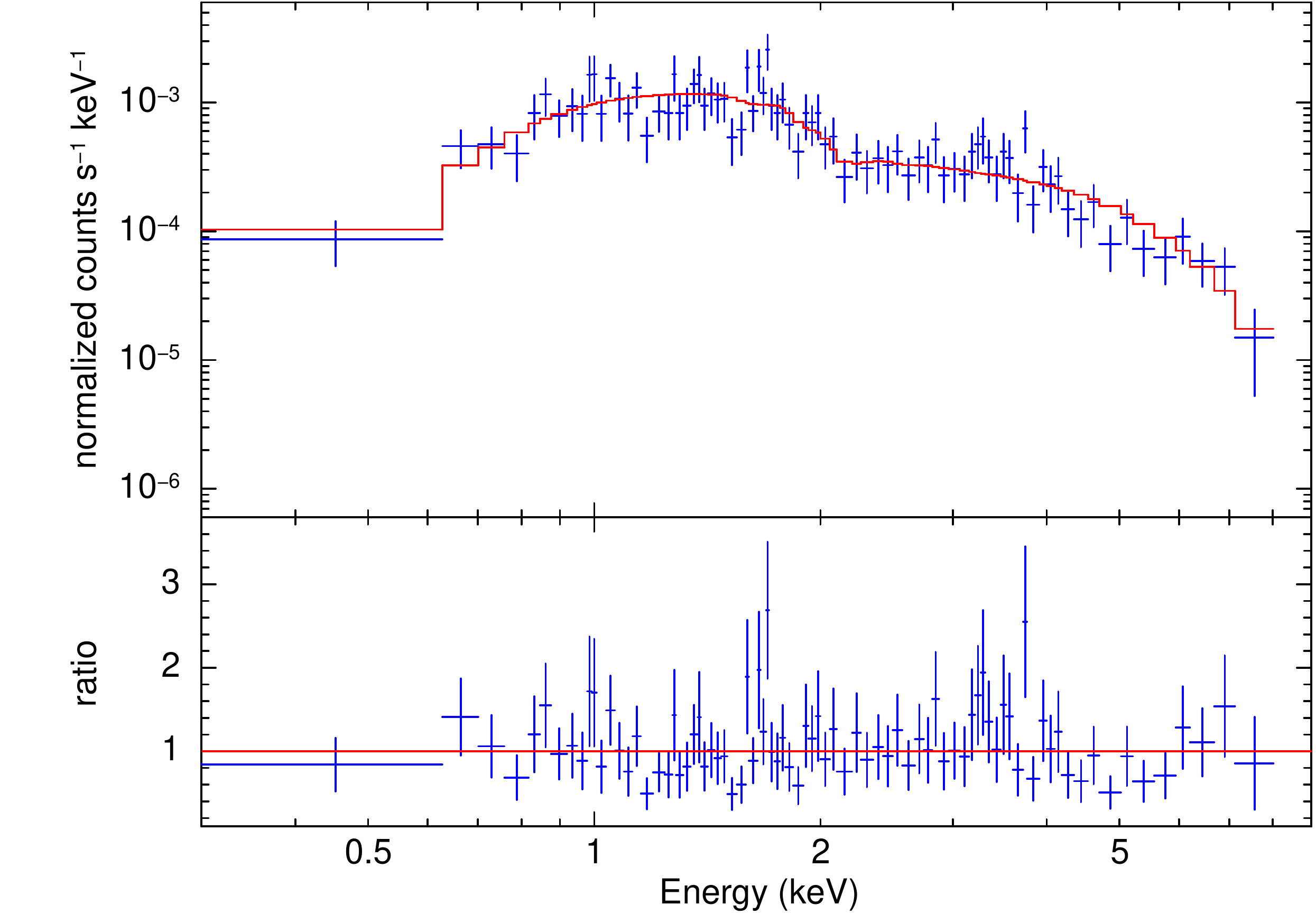}
 \caption{Combined spectrum of X-1 with best-fitting power-law model and data/model ratios. See Table 4 for the fit parameters. The spectrum was fitted with the Cash statistics; the datapoints have been rebinned to a signal-to-noise ratio $>$2.5 for plotting purposes only.}
  \label{ulx1}
\end{figure}

\begin{figure}
\hspace{-0.5cm}
\includegraphics[height=0.482\textwidth, angle=270]{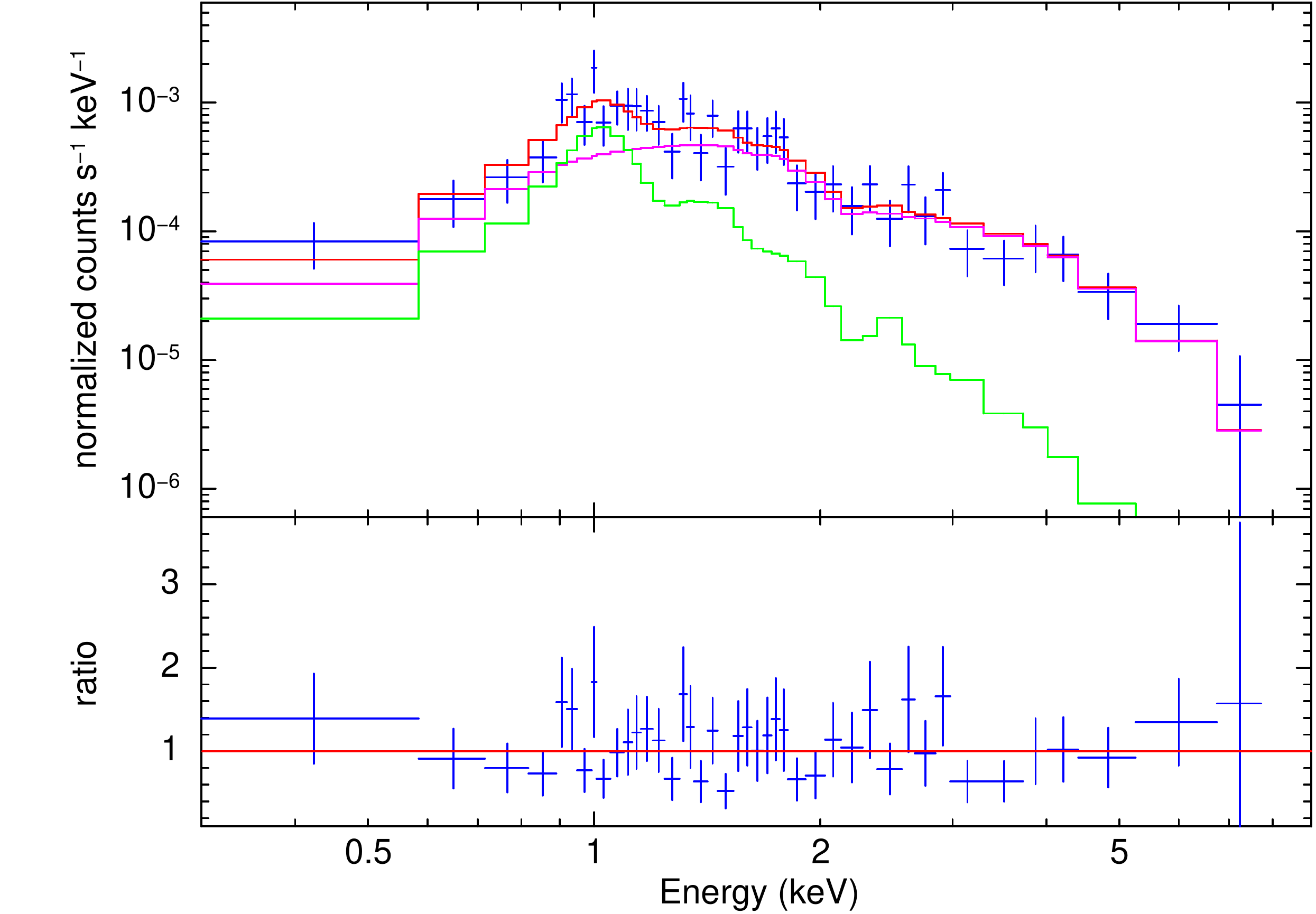}
 \caption{Combined spectrum of X-2 with best-fitting disk-blackbody plus thermal plasma model (red histogram), and data/model ratios. The magenta and green histograms are the {\it diskbb} and {\it apec} contributions, respectively. See Table 5 for the fit parameters. Datapoints were rebinned for plotting purposes.}
  \label{ulx2}
\end{figure}


\begin{figure}
\hspace{-0.5cm}
\includegraphics[height=0.482\textwidth, angle=270]{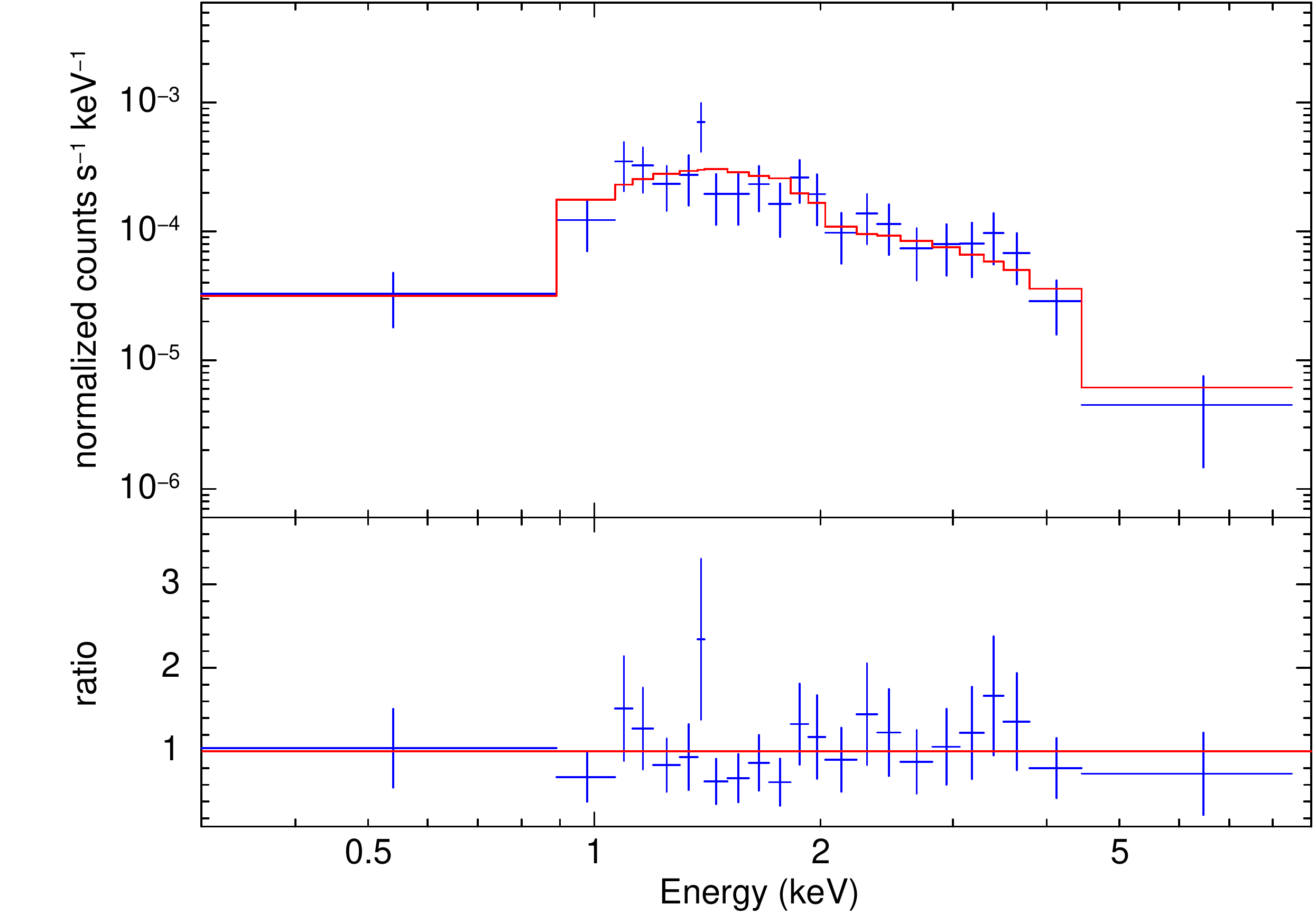}
\caption{Combined spectrum of X-3 with best-fitting disk-blackbody model and data/model ratios. See Table 6 for the fit parameters. Datapoints were rebinned for plotting purposes.}
  \label{ulx3}
\end{figure}


\begin{table}
\caption{As in Table 4, for the combined spectrum of X-3.} 
\vspace{-0.2cm}
\begin{center}  
\begin{tabular}{lc} 
 \hline 
\hline \\[-8pt]
  Model Parameters      &      Values \\
\hline\\[-8pt]
\multicolumn{2}{c}{{\it tbabs} $\times$ {\it tbabs} $\times$ {\it po}}\\
\hline\\[-8pt]
   $N_{\rm {H,int}}$   ($10^{20}$ cm$^{-2}$)   &  $ 69.1^{+48.5}_{-36.5}$ \\[4pt]
   $\Gamma$      &  $ 2.12^{+0.56}_{-0.49}$       \\[4pt] 
   $N_{\rm {po}} $ 
              &  $ 2.7^{+2.3}_{-1.1}$ \\[4pt]
   C-stat/dof     &      $81.8/111$ (0.74) \\[4pt]
   $f_{0.5-7}$ ($10^{-14}$ erg cm$^{-2}$ s$^{-1}$) 
       & $ 0.63^{+0.12}_{-0.12}$\\[4pt]
   $L_{0.3-10}$ ($10^{40}$ erg  s$^{-1}$)
      & $ 1.47^{+1.04}_{-0.38}$\\[2pt]
   \hline\\[-8pt]
\multicolumn{2}{c}{{\it tbabs} $\times$ {\it tbabs} $\times$ {\it diskbb}}\\
\hline\\[-8pt]
$N_{\rm {H,int}}$   ($10^{20}$ cm$^{-2}$)   &  $ 33.5^{+31.7}_{-24.0}$  \\[4pt]
   $kT_{\rm in}$ (keV)     &  $ 1.15^{+0.39}_{-0.24}$\\[4pt] 
   $N_{\rm {dbb}} $  ($10^{-4}$ km$^2$) &  $ 2.20^{+4.02}_{-1.49}$\\[4pt]
   $R_{\rm {in}}\sqrt{\cos \theta} $  (km) & $168^{+114}_{-73}$\\[4pt]
   C-stat/dof     &   $77.4/111$ (0.70)\\[4pt]
   $f_{0.5-7}$ ($10^{-14}$ erg cm$^{-2}$ s$^{-1}$) 
       & $ 0.59^{+0.12}_{-0.12}$\\[4pt]
   $L_{0.3-10}$ ($10^{40}$ erg  s$^{-1}$) 
      & $ 0.81^{+0.17}_{-0.14}$\\[2pt]
      \hline
\vspace{-0.5cm}
\end{tabular}
\end{center}
\end{table}

\section{Discussion}

The original purpose of our work was to investigate the claimed cloud of hot gas in the disk of NGC\,1232. We found that such cloud does not exist. Thus, there is no X-ray evidence of a recent dwarf galaxy interaction. Any alleged peculiarity in the star formation density \citep{araujo18} in that region must be purely coincidental. The absence of such disk perturbation is consistent with the recent results of \cite{lima-costa20}. 

Our result is also a reminder that researchers need to be very careful when they apply adaptive smoothing techniques to X-ray images of nearby galaxies, as such smoothing may generate fake regions of diffuse emission.

Instead, we discovered three previously overlooked ULXs in the apparent companion galaxy NGC\,1232A. All three ULXs reach $\approx$10$^{40}$ erg s$^{-1}$ during at least some of the {\it Chandra} observations; X-1 and X-2 are persistently above 10$^{40}$ erg s$^{-1}$ and peak at $\approx$7 $\times 10^{40}$ erg s$^{-1}$ and $\approx$5 $\times 10^{40}$ erg s$^{-1}$, respectively.

Finding three ULXs of such luminosity in a relatively small, isolated field galaxy is unusual: on average, we expect only $\approx$1 ULX at a luminosity $\gtrsim$10$^{40}$ for a star formation rate of $\approx$10 $M_{\odot}$ yr$^{-1}$ \citep{mineo12}. Even at 93 Mpc, NGC\,1232A is a small galaxy. Its absolute $K$-band magnitude $M_K \approx -21.0$ mag (HyperLEDA) corresponds to a stellar mass $M_{\ast} \approx 2.6 \times 10^9 M_{\odot}$, where we used a conversion factor $M/L_K \approx 0.50$  \citep{bell01}.
This is approximately the same stellar mass as the Large Magellanic Cloud \citep{vandermarel06}, but NGC\,1232A has a more regular structure.   

A better comparison is the nearby ($d \approx 4.3$ Mpc) late-type barred spiral NGC\,1313 \citep{silva-villa12}, which has two persistent-but-variable ULXs with $L_{\rm X} \gtrsim 10^{40}$ erg s$^{-1}$ \citep{sathyaprakash20,pinto20,pinto16,bachetti13,grise08}. 
The two galaxies have similar morphology and (within a factor of 2) similar stellar mass (cf.~$M_K \approx -20.4$ mag for NGC\,1313) and blue luminosity ($M_B \approx -19.7$ mag for NGC\,1232A; $M_B \approx -19.1$ mag for NGC\,1313). NGC\,1232A is larger, with a D25 major axis of $\approx 23$ kpc, compared with $\approx$13 kpc for NGC\,1313. Based on the star formation rate of $\approx$0.7 $M_{\odot}$ yr$^{-1}$ in NGC\,1313 \citep{hadfield07}, we plausibly expect a rate of $\sim$1 $M_{\odot}$ yr$^{-1}$ in NGC\,1232A. There are no published studies of the SFR in NGC\,1232A, and its recession speed is high enough that its H$\alpha$ emission falls outside the standard narrow-band H$\alpha$ filters. A few mid-infrared measurements at various locations in NGC\,1232A are available from  in the {\it Wide-field Infrared Survey Explorer} ({\it WISE}; \citealt{wright10,cutri12}). A luminosity-weighted average of the {\it WISE} colors gives us $W1-W2 \approx 0.16$ mag, $W2-W3 \approx 3.48$ mag, where $W1$, $W2$ and $W3$ are the magnitudes in the 3.4-$\mu$m, 4.6-$\mu$m and 12-$\mu$m bands, respectively. This places NGC\,1232A right on the ``star-formation sequence'' of normal disk galaxies \citep{jarrett19}, in the ``active disk'' sector of the {\it WISE} colour plane. (As a comparison, the {\it WISE} colours of our ``sister'' galaxy NGC\,1313 are $W1-W2 \approx 0.05$ mag, $W2-W3 \approx 3.30$ mag, very close to NGC\,1232A along the same sequence). Nearby disk galaxies along the mid-infrared star-formation sequence and with a stellar mass similar to that of NGC\,1232A have indeed a star formation rate between $\approx0.5 M_{\odot}$ yr$^{-1}$ and $\approx$1 $M_{\odot}$ yr$^{-1}$ \citep{jarrett19}.

One explanation for the overdensity of ULXs in small star-forming galaxies is their lower-metallicity environment
\citep{mapelli10,prestwich13,brorby14,lehmer20}. 
The only metallicity study of NGC\,1232A in the literature \citep{vanzee99} is based on a long-slit spectrum taken with the Palomar 5-m telescope. The slit covers the nuclear region and a bright H {\footnotesize{II}} region $\approx$19$^{\prime\prime}$ to the south. The latter is close to but $\approx$2$^{\prime\prime}$--3$^{\prime\prime}$ to the west of X-2, which is not on the slit.  
\cite{vanzee99} find an oxygen abundance of $12 + \log({\mathrm{O}}/{\mathrm{H}}) = 8.21 \pm 0.10$ at the spiral arm location, much lower than in the nuclear region ($12 + \log({\mathrm{O}}/{\mathrm{H}}) = 8.79 \pm 0.10$).  As a comparison, NGC\,1313 has an oxygen abundance of $12 + \log({\mathrm{O}}/{\mathrm{H}}) \approx 8.1$ at the same relative fraction of its D25 radius \citep{pilyugin14}. The mean values for the Large and Small Magellanic Clouds are $12 + \log({\mathrm{O}}/{\mathrm{H}}) = 8.35 \pm 0.03$ and $12 + \log({\mathrm{O}}/{\mathrm{H}}) = 8.03 \pm 0.03$, respectively \citep{cipriano17}. Thus, NGC\,1232A is among the most metal-poor disk galaxies, but not exceptionally so. \cite{lehmer20} showed that galaxies with low metal abundances formed an increased number of high-mass X-ray binaries and ULXs at luminosities $L_{\rm X} \gtrsim 10^{38}$ erg s$^{-1}$; at a metallicity of $12 + \log({\mathrm{O}}/{\mathrm{H}}) \approx 8.2$, the probability of finding a ULX with a 0.5--8 keV luminosity $>$10$^{40}$ erg s$^{-1}$ is $\approx$0.2 for a star formation rate of 1 $M_{\odot}$ yr$^{-1}$. This is about three times higher than the ULX occurrence rate observed in solar-metallicity galaxies \citep{lehmer20}, but it is still far too low to explain the three ULXs seen in NGC\,1232A. If the observed number of ULXs for an individual galaxy is drawn from a Poisson distribution with the expectation value set by the prediction from \cite{lehmer20}, the presence of three bright ULXs in NGC\,1232 corresponds to a probability of $\approx$10$^{-3}$. This appears an impossibly low probability; however, observations and simulations suggest that the number density of galaxies with stellar mass $>$10$^9 M_{\odot}$ in the local universe is $\sim$0.02 (Mpc)$^{-3}$ \citep{karachentsev13,torrey15,conselice16,bernardi17}, corresponding to $\sim$10$^5$ galaxies like NGC\,1232 or bigger within 100 Mpc. On the other hand, we also have to consider that only a few 100 nearby galaxies have been observed by {\it Chandra}: so, within the observed sample, NGC\,1232A must be a very rare galaxy, which deserves further multiband studies of its star formation properties and compact object activity.




\section*{Acknowledgements}
RS acknowledges hospitality at Sun Yat Sen University in Zhuhai, during part of this work. We thank Tom Jarrett for his explanations about star formation rates from {\it WISE} bands, and Alister Graham for comments about disk galaxy structure. We also appreciated the useful comments and suggestions from the referee. This publication makes use of data products from the Wide-field Infrared Survey Explorer, which is a joint project of the University of California, Los Angeles, and the Jet Propulsion Laboratory/California Institute of Technology, funded by the National Aeronautics and Space Administration. 

\section*{Data Availability}
The raw {\it Chandra} datasets used for this work are all available for download from their respective public archives. The reduced data can be provided upon request.


\bsp	
\label{lastpage}

\end{document}